# Spin switches for compact implementation of neuron and synapse

Vinh Quang Diep[1,a)], Brian Sutton[1], Behtash Behin-Aein[2] and Supriyo Datta[1]
[1] *School of Electrical and Computer Engineering, Purdue University, West Lafayette, IN, 47907 USA.*
[2] *GLOBALFOUNDRIES Inc., Sunnyvale, California 94085, USA.*

## Abstract:

Nanomagnets driven by spin currents provide a natural implementation for a neuron and a synapse: currents allow convenient summation of multiple inputs, while the magnet provides the threshold function. The objective of this paper is to explore the possibility of a hardware neural network (HNN) implementation using a spin switch (SS) as its basic building block. SS is a recently proposed device based on established technology with a transistor-like gain and input-output isolation. This allows neural networks to be constructed with purely passive interconnections without intervening clocks or amplifiers. The weights for the neural network are conveniently adjusted through analog voltages that can be stored in a non-volatile manner in an underlying CMOS layer using a floating gate low dropout voltage regulator. The operation of a multi-layer SS neural network designed for character recognition is demonstrated using a standard simulation model based on coupled Landau-Lifshitz-Gilbert (LLG) equations, one for each magnet in the network.

The standard building block[1] for neural networks (Fig.1a) consists of 1) a synapse that multiplies a number of input signals $x_i$ with approriate weights $w_i$ and 2) a neuron that sums all the weighted inputs together with a fixed bias $b$ to produce an output $y$ determined by some nonlinear function "$f$"

$$y = f\left(\sum_i w_i x_i + b\right) \qquad (1)$$

It is well established that multilayer neural networks obtained by interconnecting building blocks of this form can be designed to implement useful functions and powerful algorithms have been developed for choosing the weights $w_i$ and the bias $b_i$ so as to implement a desired overall input-output functionality.

Most neural networks are usually implemented through software although it is recognized that hardware implementations could potentially[2,3] lead to significant speed, power improvements and massively parallel computation [4]. Different proposals based on spin torque devices, domain wall motions and memristors were previously proposed to implement neurons or/and synapses [5-8]. The objective of this paper is to demonstrate the feasibility of implementing a hardware neural network (HNN) using a spin switch (SS)[9] as the basic building block, by presenting a concrete implementation of a SS neural network for character recognition and establishing its operation through direct simulation using experimentally benchmarked models for SS devices. SS has a gain that gives it a transistor-like character allowing multiple units to be interconnected without intervening CMOS circuitry for clocks or amplifiers: one spin switch can directly drive the next one like ordinary transistors.

Fig. 1b shows a schematic representation and detailed structure of a building block implemented with a spin switch. It consists of a *Write (W)* unit and a *Read (R)* unit with free magnetic layers $x_i, x_i{}'$ or $(y, y')$ that are dipole coupled ensuring that the two are always anti-parallel $x_i = -x_i'$. This configuration allows information to propagate from the *Write*

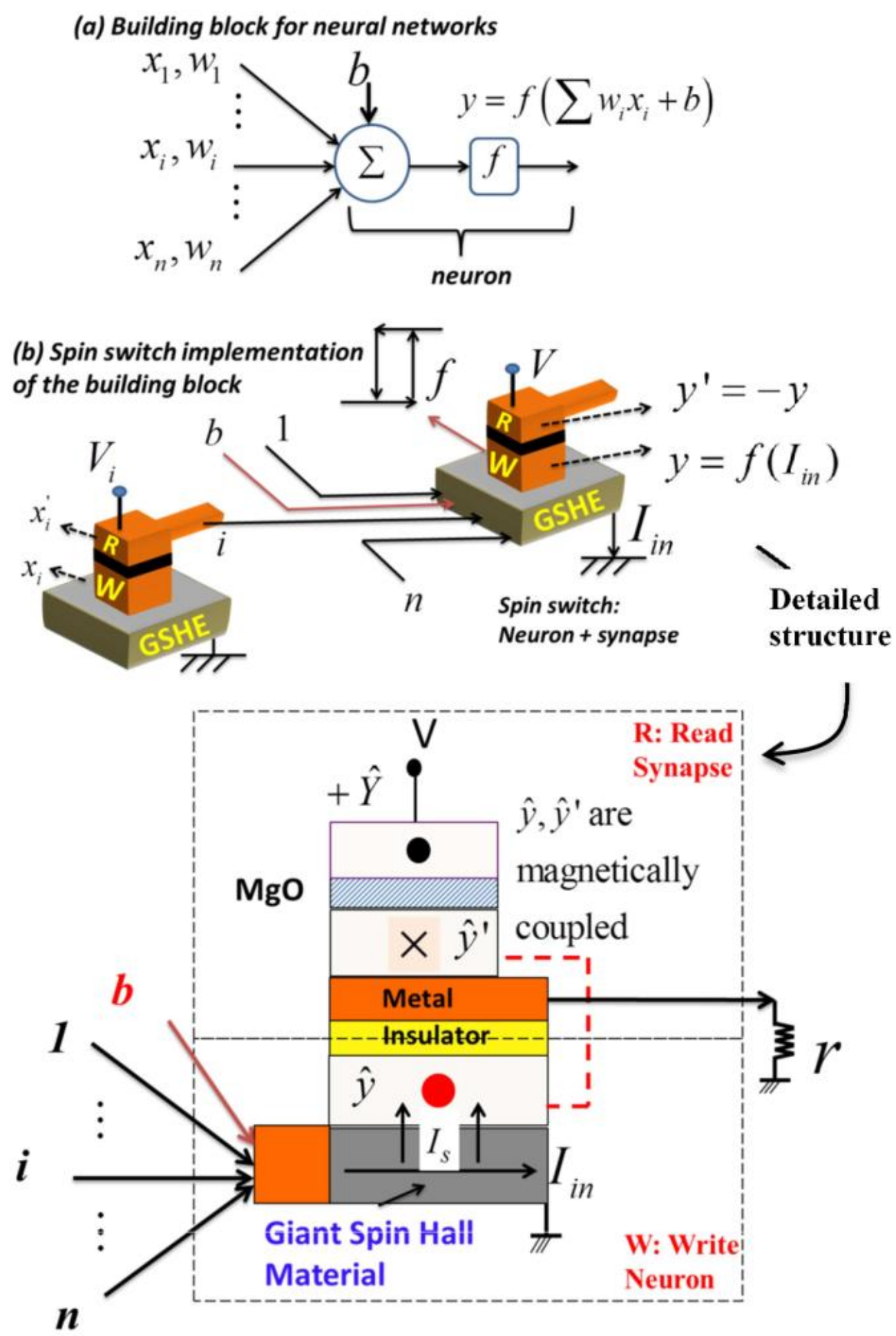

**FIG. 1 a)** Standard model for the basic building block of neural networks: A neuron sums the incoming signal $\boldsymbol{x_i}$ with weights $\boldsymbol{w_i}$ and generates the output according to the activation function $\boldsymbol{f}$ as expressed by Eq.(1). **b)** Spin switches can be used to provide a compact implementation of this building block as evident from comparing Eq.(3) with Eq.(1).

to the *Read* unit of a spin switch while maintaining their electrical isolation. As we will see, the *Write* unit functions as a neuron which performs the summation and threshold

a) vdiep@purdue.edu

function $f$ while the *Read* unit functions as a synapse which provides the weighted output $w_i x_i$.
The *Read* unit consists of one MTJ whose conductance is $G_i = g_i/2(1+P_i\hat{Y}_i.\hat{y}_i') = g_i/2(1-P_i\hat{Y}_i.\hat{y}_i)$ where $P_i = (G_P - G_{AP})/g_i$; $g_i = G_P + G_{AP}$; $\hat{Y}_i$ (or $\hat{X}$) is the magnetization of fixed magnet and $\hat{y}_i$ (or $\hat{x}'$) is the magnetization of free magnet.

The *Write* unit consists of a giant spin Hall effect (GSHE) metal like Ta[10] or tungsten[11] and a free magnet $\hat{x}_i$ (or $\hat{y}$) which generates a spin current $I_s$ when driven by a charge current $I_{in}$: $I_s = \beta I_{in}$, $\beta = \theta_{SH}(A_s/A)$ and $\theta_{SH}$ being the spin-Hall angle. $A_s$ and $A$ are the cross-sectional areas for the spin and charge currents respectively.

The *Read* unit functions as a synapse whose output current is weighted by $\sim V_i G_i$ while the *Write* unit functions as a neuron (see Fig. 1b): it sums weighted inputs from the R units of many spin switches labeled $1, \dots, i, \dots, n$ giving rise to a net current

$$I_{in} = \sum_i V_i \frac{G_i(\hat{x}_i)}{1+G_{total}\, r} + I_b, \text{ where } G_{total} \equiv \sum_i G_i \qquad (2)$$

($r$: resistance of the GSHE metal) and performs a hysteretic threshold function $f$ of the form sketched in the figure to produce the output:

$$y = f(I_{in}) = f\left(\sum_i \frac{V_i\, g_i}{2(1+G_{total}\, r)}(1-P_i \hat{X}_i.\hat{x}_i) + I_b\right) \qquad (3)$$

Comparison with Eq.(1) suggests that spin switches could provide a possible building block for HNN using $V_i$ to implement the desired weights $w_i = V_i g_i/2(1+G_{tot} r)$. However, the details are not obvious especially since the function $f$ is hysteretic rather than the usual single-valued function.

In estimating the threshold current/voltage needed to drive the (next) *Write* unit, we note that a spin current of $2I_{sc}$ (critical spin current required to switch the magnet) will be needed to flip the *Write* magnet since it couples to the magnet of the *Read* unit. Hence:

$$I_{th} \geq \frac{2I_{sc}}{\beta} = \frac{2I_{sc}}{\theta_H (A_s/A)} \qquad (4)$$

As a result, the threshold voltage

$$V_{th} \geq \frac{2I_{th}}{g_i(1-P)}(1+G_i\, r) \qquad (5)$$

Using parameters described in Ref. [9], we have $I_{sc} = 160\mu A$. If $t_{gshe} = 2nm$ and the Hall angle reported experimentally $\theta_{SH} = 0.3$ for tungsten, the threshold current $I_{th} \sim 30\mu A$.

Also assuming TMR of 135% for the MTJ and resistance-area product of $A_s/G_P = 4.3\Omega\mu m^2$ [12], we have $P = 0.4$ and $g_i = (1.1k\Omega)^{-1}$ for $A_s = 80nm \times 30nm$. If we chose other GSHE material [13,14] with comparable spin hall angle but with low resistivity, then we can assume $G_i r \ll 1$. This will give us $V_{th} \sim 100mV$.

The power consumption of SS can be estimated as $P = I_{th} V_{th} \sim few\ \mu W$ along with the switching time of magnet as $\tau \approx 1ns$ results in energy consumption for the SS $E \approx few\ fJ$ per switch. We should mention, however, that from the point of view of energy and delay, spin switches (or in general beyond-CMOS devices[15]) based on established technology are inferior to a single CMOS transistor, but may still look attractive compared to CMOS based neurons[16], due to the compact multi-functionality provided by the SS (note that, the area of a SS is roughly the area of magnet in the *Read* or *Write* unit which is typically $\approx 100nm^2$). Moreover, the switching energy of a single SS neuron could be lowered significantly as new phenomena are discovered and developed (see for example[17–19]).

*Fan-out:* An important attribute of the SS is the possibility of large fan-out, whereby the output from one spin switch can be used to drive hundreds of other spin switches thus allowing large interconnectivity which is important for implementing neural network functions.

Fig.2 shows a SS neuron with multiple outputs where the voltage at each R unit represents the synaptic weight of this neuron in connection with the other neurons. Since each R unit has its own independent power supply, this neuron should be able to drive a large number of outputs. Note that the interconnections between neurons do not require GSHE material. They could be low resistance material (copper wires) with no increase in energy consumption. The only limitation arises from the threshold current/voltage needed to drive such a "big" neuron. Eq. (4) shows that the threshold current is proportional to the area A of GSHE, but independent of its length. Of course the resistance 'r' of the GSHE increases with length but this has a minimal effect on the threshold voltage as long as the factor $(1 + G_i r)$ is not excessive. For example, if $L_{ghse}/L_{FM} = 100$, $W_{ghse}/W_{FM} = 2$, $\rho_{gshe} = 10^{-7}\Omega m$ we have $(1 + G_i r) \approx 3$.

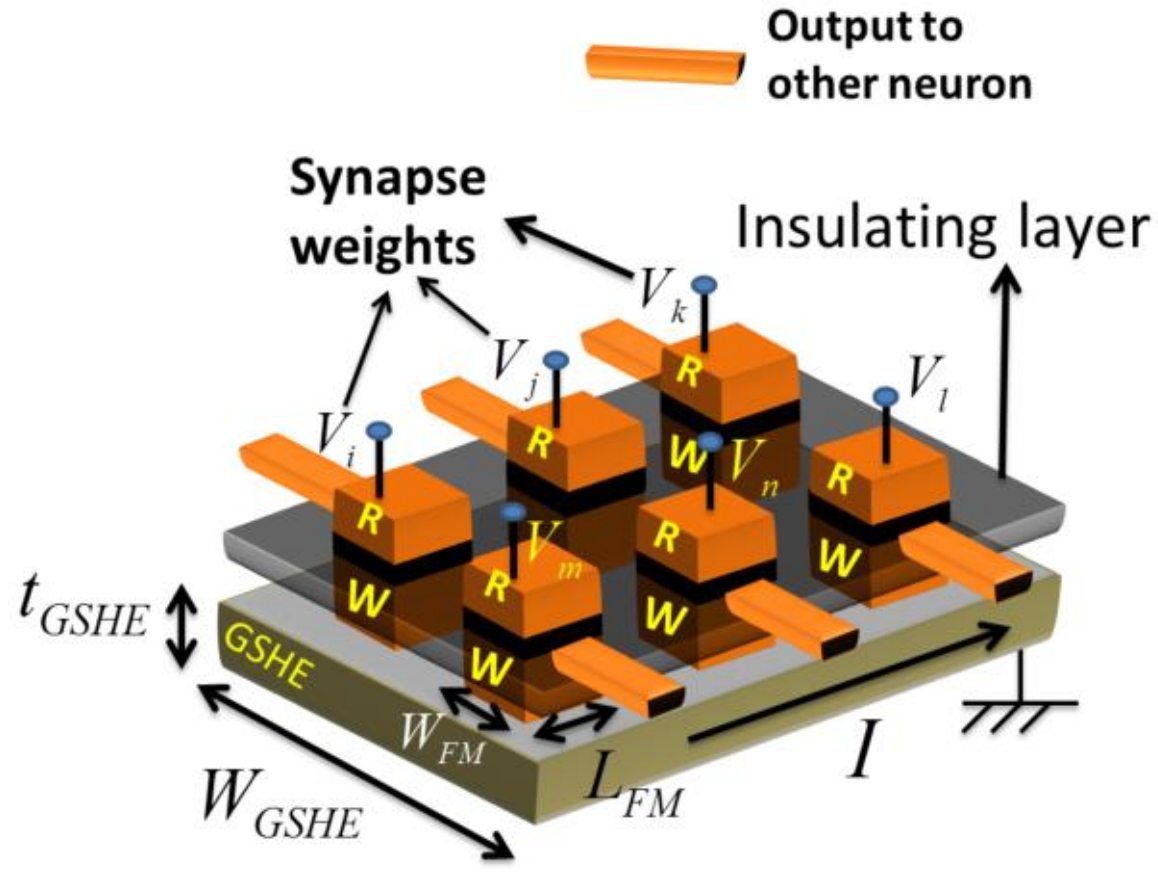

**FIG.2** A composite spin switch neuron with built-in synapses. The magnetization of any R unit represents the state of the neuron. The voltages at each R units represent the weight of synapses connecting to other neurons.

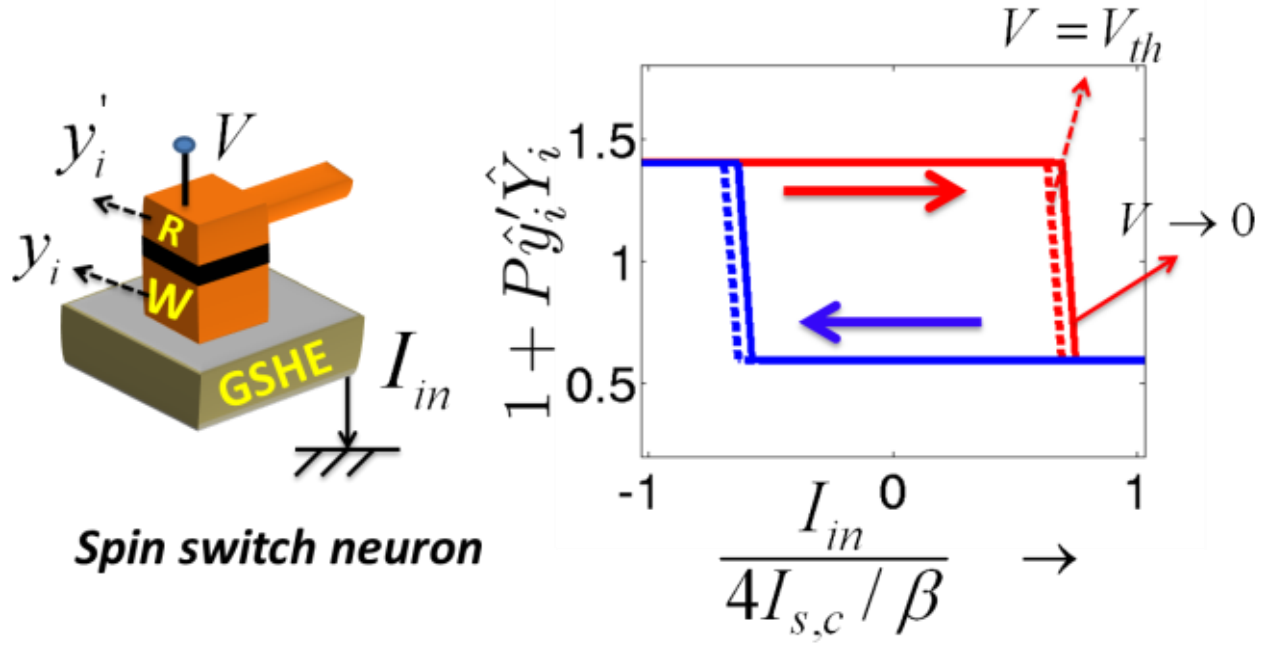

**FIG. 3**: Spin switch switching characteristic (also the activation function of SS): $\mathbf{V=0}$ (solid line) and $\mathbf{V=V_{th}}$ (dash line).

*In designing a neural network (NN),* we focus on the multi-layer feed-forward NN with training done off-line by smoothing the SS hysteresis function to make it differentiable and adapting the back-propagation method[20]. Other techniques such as the weight perturbation method[21] or the extreme learning machine[22] might also be suitable especially for on-chip training. The key difference with standard neural network design is the hysteresis in the threshold function $f$ which makes the overall design more robust, but requires an extra condition during the training: the currents arriving at each neuron have to be above threshold [23].

To model SS devices, it's notice that each SS with a W/R pair requires a pair of LLG equations for the magnet pair $\hat{x}_i$ (or $\hat{y}_i$) and $\hat{x}_i'$ ($\hat{y}_i'$) (see [9] for more detail).

$$\frac{1+\alpha'^2}{1+\alpha^2}\frac{d\hat{x}_i'}{d\tau} = -\hat{x}_i'\times\vec{h}_i' - \alpha\hat{x}_i'\times\hat{x}_i'\times\vec{h}_i' - \hat{x}_i'\times\hat{x}_i'\times\vec{i}_{si}' + \alpha\hat{x}_i'\times\vec{i}_{si}' \qquad (6)$$

$$\frac{d\hat{x}_i}{d\tau} = -\hat{x}_i\times\vec{h} - \alpha\hat{x}_i\times\hat{x}_i\times\vec{h} - \hat{x}_i\times\hat{x}_i\times\vec{i}_{si} + \alpha\hat{x}_i\times\vec{i}_{si} \qquad (7)$$

where

$$\tau = (\gamma\mu_0 H_k t)/(1+\alpha^2); \vec{h} = \vec{H}/H_k; \vec{h}' = \vec{H}'/H_k$$

The total field

$$\vec{H}' = H_k' x_i'(z)\hat{z} - H_d' x_i'(y)\hat{y} - H_b\hat{x}_i$$

$$\vec{H} = H_k x_i(z)\hat{z} - H_d x_i(y)\hat{y} - H_f\hat{x}_i'$$

includes the easy axis field $(H_k', H_k)$, demagnetization field $(H_d', H_d)$, and dipolar field $(H_b, H_f)$. The dimensionless spin currents in Eq. (6) and Eq. (7) are given by

$$\vec{i}_{si} = \frac{\vec{I}_{si}}{(2q/\hbar)\mu_0 H_k M_s \Omega} \quad ; \quad \vec{i}_{si}' = \frac{\vec{I}_{si}'}{(2q/\hbar)\mu_0 H_k M_s \Omega}$$

where $M_s', M_s$ are the saturation magnetizations and $\Omega$ is the volume of the magnet. The following parameters are used for all magnets:

$$\alpha = 0.01; H_d = 50H_k, H_f = H_b = H_k = 0.02T/\mu_0$$

$$\mu_0 M_s = 1T;\ \Omega = 80nm\times 100nm\times 1.6nm$$

The spin currents are obtained by summing the inputs from the preceding *Read* units.

$$\vec{I}_{si} = \beta\left(\sum_j V_j \frac{G_j(\hat{x}_j)}{1+r_i G_{\text{total}}} + I_{bi}\right)\hat{z};\quad \vec{I}_{si}' = -P_i V_i G_i \hat{z}$$

where $V_i$ are the voltages and $I_{bi}$ are bias currents applied at preceding Read units.

Fig. 3 shows the input-output characteristic of a SS device. Note that at $V=0$, the switching currents are symmetric but there is a *minor* shift when $V=V_{th}$. This is due to a (small) spin current $I_s'$ injected by the *Read* unit making it easier to switch from +1 to -1 than to switch from -1 to +1. This shift is relatively minor ensuring the key property of *directionality:* the state of the magnet (neuron) is largely determined by the input and not the output current.

*Neural network for character recognition*: Now we consider an example of a SS neural network designed for character recognition: it recognizes 8 letters A, B, C…H represented by $7\times 5$ matrices or vectors having 35 components (zero or one). Fig. 4a shows the layout of the SS neural network for this pattern recognition with 35 neurons in the input layer, 6 neurons in the hidden and 3 neurons in the output layer (labeling from 1 to 9). Hence each neuron in the input layer has six synapses while each neuron in the hidden has 3 synapses.

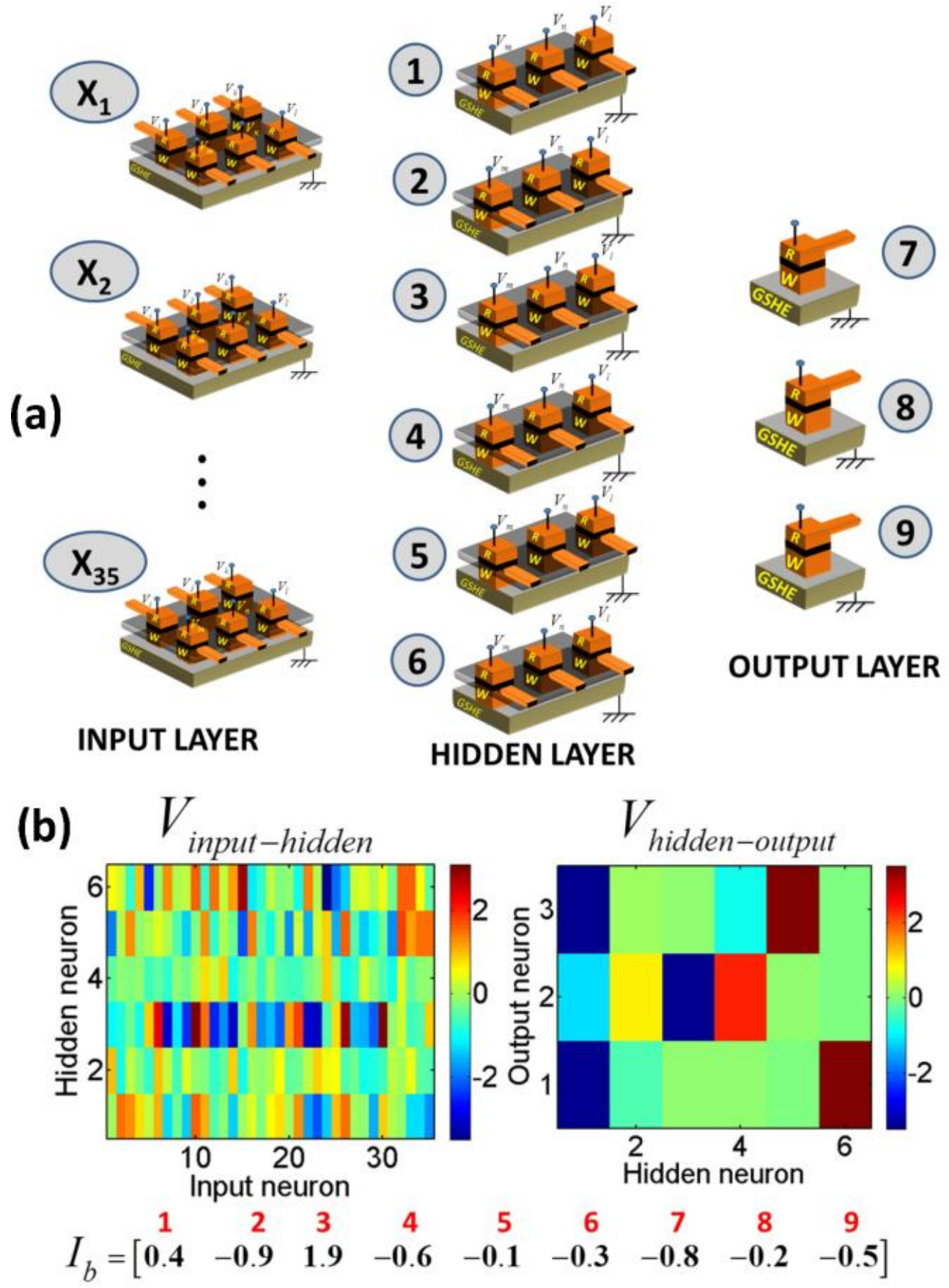

**FIG.4 a)** Implementation of spin switch neural network for character recognition: for simplicity the connections (copper wires) between layers are not shown. **b)** Matrices of interconnection weight voltages between input and hidden layers $(\mathbf{35\times 6})$ and between hidden and output layers $(\mathbf{6\times 3})$ shown by color scale.

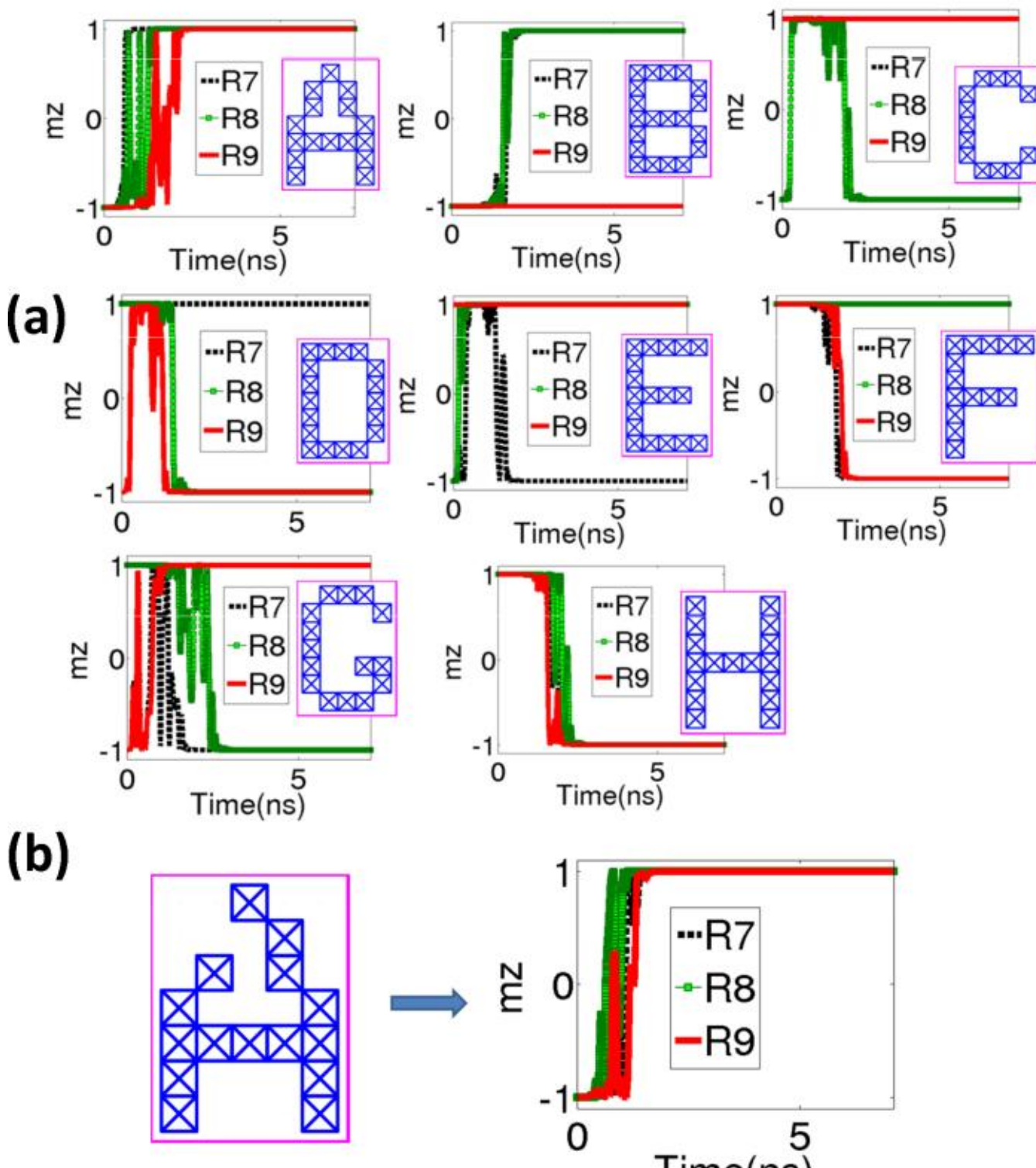

**FIG. 5 a)** LLG simulations for the outputs layer of the SS neural network described in Fig.5a in response to eight input characters: A, B, C…H. Here R7, R8 and R8 are the magnetizations of magnets in the R units of neuron 7, 8 and 9 respectively. **b)** The network still recognizes a letter that is not in the training set (a random defect). We assume all *Read* units have the same polarization and conductance: $\mathbf{P_i = 0.4, g_i = (1.1k\Omega)^{-1}; \;\; \beta = 12}$

The voltages $V$ applied at *Read* units represent the weight of synapse in connecting with other neurons while the $I_b$ represents the bias current applied at each neuron. Both voltages and bias currents obtained through the training are shown in Fig. 4b in the form of $(35 \times 6)$, $(6 \times 3)$ matrices and $(1 \times 9)$ vector. They are all normalized to their threshold values.

Fig. 5a shows simulation results obtained from solving a set of 88 coupled LLG (70 for the input, 12 for the hidden and 6 for the output layer) equations whose solutions are the dynamics of the magnets. For a neuron with many synapses, we assume all the W/R units are identical so that the whole neuron can be modeled as a single W/R unit. Fig. 5a illustrates the character recognition function: If the input is A, the output is 000 which can be translated to $\bar{1}\bar{1}\bar{1}$ for magnetizations of W magnets (note $\bar{1}$ means $m_z = -1$). For magnets in the R units the result is 111. It is interesting to notice that, even there is a (random) defect in the letter, the network is still able to recognize it as shown in Fig. 5b.

It's also notice that the layout for character recognition in Fig. 4a is quite robust to small variations of voltages and bias currents (see also XOR example in [23]). Since currents arriving at neurons are always above threshold, the switching characteristics will not be affected by thermal noise.

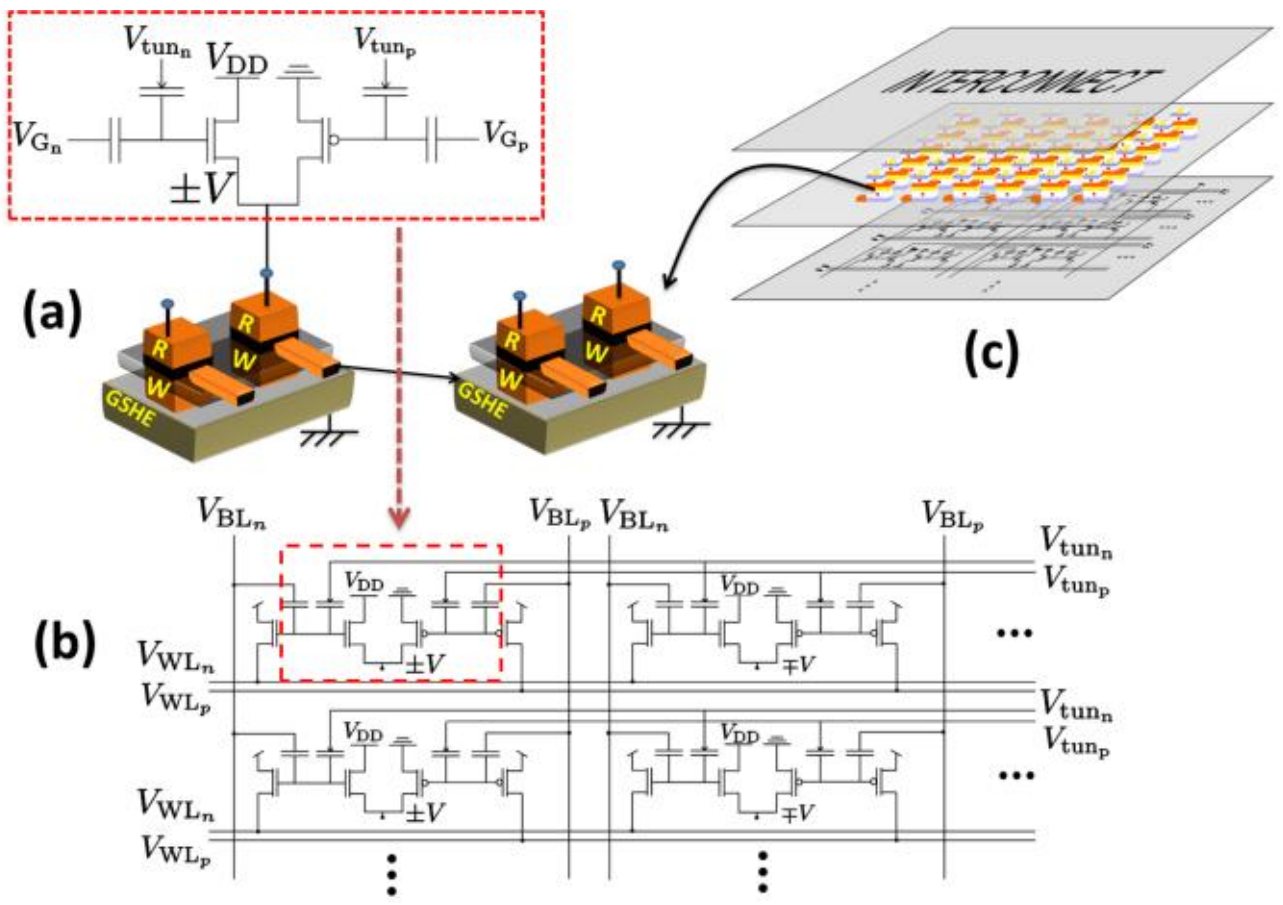

**FIG. 6 Possible way of supplying/storing long term voltage a)** The weights (+/- V) of the MTJ stack are provided through the use of tunable complementary floating-gate low dropout voltage regulators. Channel Hot Electron Injection and Fowler-Nordheim tunneling modulated by $V_G, V_{tun}$ can be used to adjust the charge trapped on the floating gate. In turn, those charges adjust the threshold of the nFET/pFET devices giving $+/-V$ at the MTJ of the **R** unit. **b)** Each floating-gate may be formed in a bit-addressable array for individual nFET/pFET threshold tuning. **c)** The spin-switch layer is formed in the metallization above the silicon layer along with the interconnect for the neural network.

Finally we would like to discuss a possible way of storing and adjusting the voltage applied at R units. The use of floating-gate(FG) transistors has been proposed as a mechanism to store tunable analog voltages which can be used for synapse weights[24,25]. Controlled amounts of charge can be injected and removed from the gate to create a spectrum of gate voltages which is important in the realm of neural networks.

Fig. 6 shows a possible means to store/adjust the voltages applied at the synapse *Read* unit MTJs using a FG low drop-out voltage regulator[26]. As depicted, programmatic control of an individual FG charge can be accomplished through an addressable array that selectively controls $(V_{\mathrm{tun,n}}, V_{\mathrm{Gn}})$ and $(V_{\mathrm{tun,p}}, V_{\mathrm{Gp}})$. The tunneling voltage can be used to provide a global erase of stored values via Fowler-Nordheim tunneling while the respective gate voltages can be used to inject charge onto the FG with Channel Hot Electron Injection. This charge control in turn modulates the threshold voltage of the nFET/pFET and hence the regulated voltage. As a result, the nFET/pFET can be programmed to provide the desired synapse weights through an active learning process. Once programmed, these FG transistors will retain the voltage for an extended period of time due to their non-volatility.

*In summary,* we have demonstrated the possibility of a hardware neural network implementation using a spin switch (SS) as its basic building block. SS is a recently proposed device based on established technology having a transistor-like gain and input-output isolation that allows large circuits to be constructed without intervening clocks or amplifiers. The SS neuron-synapse used in the present paper differs from the SS proposed in [9] in the sense that there is only one MTJ stack at the Read unit. Among the three components comprising the SS, so far the GSH material and MTJ stack have been experimentally established. The dipolar coupling has also been well studied in the context of nano-magnet logic[27]. But for SS applications, one needs to show the dipolar coupling between magnets having thicknesses $2-4\mathrm{nm}$.

We have shown that a SS can be used to build neuron capable of performing summation, multiplication and an activation function which normally requires extra circuitry in other hardware implementations. The SS neurons occupy area much less than $\mu m^2$, consume femtojoules per switch and operate at room temperature. The weights for the neural network are conveniently adjusted through analog voltages that can be stored in a non-volatile manner in an underlying CMOS layer using a floating gate low dropout voltage regulator. The operation of a multi-layer SS neural network designed for character recognition is demonstrated using a standard simulation model based on coupled Landau-Lifshitz-Gilbert (LLG) equations, one for each magnet in the network.

It is a pleasure to thank Paul E. Hasler for useful discussion about floating gate low dropout regulators. This work is supported by the Center for Science of Information (CSoI), an NSF Science and Technology Center, under grant agreement CCF-0939370.